\documentclass[final,1p,times]{elsarticle}
\usepackage{graphicx}

\usepackage[export]{adjustbox}
\usepackage{epsfig}
\usepackage{graphicx}
\usepackage{caption}
\usepackage{subcaption}
\usepackage[export]{adjustbox}
\usepackage{wrapfig}

\usepackage{amssymb}
\usepackage{amsmath}
\usepackage{algpseudocode}
\usepackage{algorithm}
\usepackage{todonotes}
\usepackage{rotating}
\usepackage{multirow}
\usepackage{array}
\usepackage{tabto}
\usepackage{color, soul}
\usepackage{url}
\usepackage[colorlinks]{hyperref}
\usepackage{cleveref}

\begin{document}
\begin{frontmatter}


\title{\#EndSARS Protest: Discourse and Mobilisation on Twitter}

\author{Bello Shehu Bello}
\author{Muhammad Abubakar Alhassan}
\author{Isa Inuwa-Dutse}

\begin{abstract}
    Using the \textit{@NGRPresident} Twitter handle, the Government of Nigeria issued a special directive banning Special Anti-Robbery Squad (SARS) with immediate effect. The SARS is a special police unit under the Nigeria Police Force tasked with the responsibility of fighting violent crimes. However, the unit has been accused of waves of human rights abuse across the nation. According to a report by Amnesty International, between January 2017 and May 2020, 82 cases of police brutality have been committed. This has led to one of the major protests demanding more measures to be taken. The \textit{\#EndSARS} hashtag was widely used by the protesters to amplify their messages and reach out to wider communities on Twitter. 
    In this study, we present a critical analysis of how the online protest unfolded. Essentially, we examine how the protest evolves on Twitter, the nature of engagement with the protest themes, the factors influencing the protest and public perceptions about the online movement. We found that the mobilisation strategies include direct and indirect engagements with influential users, sharing direct stories and vicarious experiences. Also, there is evidence that suggests the deployment of automated accounts to promote the course of the protest. In terms of participation, over 70\% of the protest is confined within a few states in Nigeria, and the diaspora communities also lent their voices to the movement. The most active users are not those with high followership, and the majority of the protesters utilised mobile devices, accounting for 88\% to mobilise and report on the protest. Moreover, we have examined how social media users interact with the \#EndSARS movement and the response from the wider online communities. Needless to say, the themes in the online discourse are mostly about ending SARS and vicarious experiences with the police, however, there are topics around police reform and demand for regime change. We surmise that injecting a political undertone could lead to setbacks in social activism as it will likely provoke counter protest. 
\end{abstract}
\begin{keyword}
Online Activism, Online Protest, EndSARS Protest, Online Social Networks, Twitter
\end{keyword}
\end{frontmatter}

\section{Introduction} 
\label{sec:introduction}
On October 11, 2020, a special directive banning Special Anti-Robbery Squad (SARS) was issued via the Twitter handle\footnote{\url{https://twitter.com/NGRPresident/status/1315273093221318656}} of the Presidency Nigeria (see Figure~\ref{fig:SARS-ban}). The SARS is a special police unit under the Nigeria Police Force, responsible for fighting violent crimes such as robbery and kidnapping. The unit has been accused of various atrocities and human rights abuse including torture and murder across the nation. In a report compiled by Amnesty International, there are various cases of torture, ill-treatment extortion, and extrajudicial executions carried out by officers of the SARS \cite{amnesty-sars2020}. According to the report, between January 2017 and May 2020, 82 cases of police brutality have been committed. The tipping point of the protest is attributed to the period when a video of a man being killed went viral. 
This and confounding allegations of police brutality led to an intense spate of protests against the activities of SARS. Despite the Presidential directive, the protest continues because the protesters view the details of the disbandment to be insufficient. 
The dissatisfaction led to intense protests demanding reforms beyond SARS, but the Nigeria Police Force. As a result, there were many incidences of clashes between the police and protesters. The infamous Lekki bridge has been widely reported, especially online. Prior to these events, the protest has been ongoing on Twitter for a very long time. We were able to trace one of the earliest tweets about the movement (see Figure~\ref{fig:FirstENDSARSTweet}), and the hashtag \#EndSARS appears to be the trademark of the online protest to draw attention to human rights violations labelled against the SARS police unit. Many studies analysing the attitudes of the public towards the police have been published in the past \cite{brandl1994global,weitzer2004race,warren2011perceptions,kane2002social,mastrofski2002police}. Most of these studies preceded modern social networks such as Twitter\footnote{\url{https://twitter.com/}} and Facebook\footnote{\url{https://facebook.com/}}. With the arrival of online social networks, among other things, citizens are empowered to express and report their experiences with law enforcement officers. An earlier study reported how vicarious experiences result in negative perceptions about the police \cite{warren2011perceptions}. With online social networks, rich information about primary and secondary encounters with the police can be easily retrieved. This makes it easier to track public attitudes towards the activities of the police. Public dissatisfaction and demand for social justice often result in social movements with the potential of bringing drastic social reforms \cite{jones2013march}. Online platforms facilitate social activism that attracts the attention of diverse individuals. A case in point is the \#BlackLivesMatter online movement \cite{ince2017social,buggs2017dating,byrd2017vitality,haffner2019place} and the \#EndSARS protest \cite{ohia2020covid,uwazuruike2020endsars,ajisafe2021impacts,dambo2021office,ekoh2021role,iwuoha2022protests,aidonojie2022legality,dambo2022nigeria}. 
While discourse about \#EndSARS has been ongoing on Twitter, it was the early October 2020 incidence of the police shooting that triggered public anger leading to the protest \cite{amaza2020}. Of interest to this study is to explore the following aspects in relation to the online protest:   
    \begin{enumerate} 
       \item \textit{How does the protest escalate and how engaging is the protest theme?} In this question, we are interested in studying activity trends during the \#EndSARS movement.  
        \item \textit{Who are the key online players during the \#EndSARS protest?} This question is geared towards identifying and characterising the influential stakeholders during the protest. Essentially, we are interested in determining whether the promoters of the protest reside within or outside the country (the diaspora community). It has been reported that the diaspora community is a strong player in this kind of movement \cite{olabode2016veterans,dambo2021office}. This study will focus on a different approach to offer a high-level categorisation of the users based on devices used to tweet about the protest and the location information. 
       \item \textit{Is the online protest purely about \#EndSARS?} This is crucial because, at some point, the protest has transformed from opposition to police brutality to a movement for social justice and government reforms.  
       \item \textit{What are the main themes in the protest and public perception about the movement?}
       The goal of this question is to identify topical themes in the protest and how the online public responds to the protest.  
\end{enumerate}

The remaining part of the paper is structured as follows. Section~\ref{sec:related-work} presents some related work and Section~\ref{sec:methodology} offers a detailed description of the data collection and preliminary analysis. Section~\ref{sec:result-and-discussion} offers the results and discussion and Section~\ref{sec:conclusion} concludes the study and discusses some future work. 

\section{Related work}
\label{sec:related-work}
In this section, we provide some related studies within the areas of public attitudes towards police \cite{brandl1994global,weitzer2004race,warren2011perceptions,kane2002social,mastrofski2002police}, social activism \cite{miller2000geography,sewell2001space,mccaughey2003cyberactivism,jones2013march,graham2016content,ince2017social,buggs2017dating,haffner2019place}, and the \#EndSARS movement \cite{ohia2020covid,uwazuruike2020endsars,ajisafe2021impacts,dambo2021office,ekoh2021role,iwuoha2022protests,aidonojie2022legality,dambo2022nigeria}. 

\subsection{Public Attitudes Towards Police} 
There is mixed reaction and perception about the relationship between police and the public. The attitudes of citizens towards the police have been investigated long before online social media platforms. For instance, the work of \cite{brandl1994global} is focused on the satisfaction of the citizens towards the role of police in precipitating civil disorder. Many factors have been attributed to the tense relationship between police and the public. In the USA, race is considered to be one of the most salient predictors of attitudes towards the police with African Americans expressing more dissatisfaction than Whites \cite{warren2011perceptions}. Using survey data, the work of \cite{weitzer2004race} examines the public perceptions of misconduct, such as verbal abuse, excessive force, unwarranted stops, and corruption, involving police in the United States \cite{weitzer2005racially}. 
In lower-class and high-crime communities, police officers were more likely to be verbally and physically abusive towards citizens \cite{kane2002social}. In another study, citizens who initiate disrespect towards the police were more likely to be treated disrespectfully by the police officer \cite{mastrofski2002police}. Vicarious experiences also result in negative perceptions about the police \cite{warren2011perceptions}. This could be a key player during the \#EndSARS protest since some of the protesters are angry because they have an indirect negative encounter with police brutality. Online social media platforms make it easier to track public perception or satisfaction with the activities of police and mobilise for social activism.

\subsection{Social Activism} 
Social movements have been influential in framing and shaping social discourse by making it clearer for the affected individuals or communities to understand societal issues in a specific way \cite{miller2000geography,sewell2001space}. The quest for social movements have resulted in various societal reforms. For instance, the popular \textit{March on Washington} for jobs and freedom advocating for the civil and economic rights of African-Americans \cite{jones2013march}. 
With the growing interconnectivity, social activism is becoming widespread on various online social media platforms. The challenges affecting marginalised communities and movements can be brought to the limelight by leveraging online social media platforms. Through the Twitter platform, the \#BlackLivesMatter movement has resulted in effecting change and injecting uniquely black concerns and perspectives into the national discourse \cite{graham2016content}. Past studies have explored various dimensions and impact of the \#BlackLivesMatter (or BLM) online movement \cite{ince2017social,buggs2017dating,byrd2017vitality,haffner2019place}. In the work of \cite{ince2017social}, the focus is on how social media users interact with the BLM movement and how online communities influence the framing of the movement. The authors found that the discourse is centred around solidarity or counter-movement sentiments towards the movement. The manifestation of the movement on Twitter has been studied in \cite{haffner2019place}. The BLM movement has been used by daters use to explain their responses to social justice movements around race and racism in the United States \cite{buggs2017dating}. Online activism tends to be effective via social media platforms because of the rate at which a large audience gets to be aware of the issue. 

\subsubsection{Online Social Networks} 
The Internet has been instrumental for research involving minority voices \cite{kennedy2006beyond}. By enabling interconnectivity among various devices, the Internet is one of the inventions that change the world \cite{warf2018sage}. The advancement in digital social networks has provided the opportunity for huge datasets to be used to examine the actions and behaviour of users at different levels. Access to online services or the Internet via smartphones and other digital devices makes it possible to present and amplify societal challenges that will rather go unnoticed. Data from social media platforms are quite popular for various research across domains \cite{miller2010data,haffner2018spatial}. With social media, the breadth of available data sources has expanded rapidly, making it possible to add more value to the traditional modes of generating and utilising research data. The suitability and relevance of data from Twitter have been assessed for biases inherent in social media usage in social research \cite{longley2015geotemporal}. 
Today, a large proportion of the populace relies on social media for various purposes. For instance, according to \cite{sheet2018social}, around 7 out of 10 Americans use social media to connect, engage with news content, share information and entertain themselves. While young adults were among the earliest adopters, the social media user base is growing more representative of the broader population \cite{sheet2018social}. 
Social networking sites gain the highest number of users around the world in which they provide their information, thoughts and opinions through various means, such as mobile phones, laptops, and tablets \cite{kumar2014twitter}. Social media platforms such as Twitter permit instantaneous conversation and participation at any time making them suitable for bringing societal issues to the surface. Such platforms have been instrumental in reshaping social discourse and movements by attracting diverse individuals to engage in direct or indirect communication. For research purposes, Twitter is useful in both organising and protesting. Users engage using various means - posting (tweeting), reposting or sharing (retweeting), liking, etc. A hashtag or group of hashtags are used to denote a topical discourse on Twitter. Such discourse may emanate from individuals or communities of users interested in promoting or expressing dissatisfaction. Similar to the \#ENDSARS protest, hashtags have been used to speak out against the activities of the police handling the SARS initiative. 

\subsection{\#EndSARS-related Studies}
As noted earlier, in early October 2020, a video surfaced on social media of police officers struggling and the subsequent shooting of a young man. This incidence was identified to involve the SARS unit of the Nigeria Police Force\footnote{\url{https://www.npf.gov.ng/aboutus/History_Nigeria_Police.php}} and triggered a heightened spate of the \#EndSARS protest \cite{amaza2020}. Since the incident, many studies have been published around the protest theme \cite{ohia2020covid,uwazuruike2020endsars,ajisafe2021impacts,dambo2021office,ekoh2021role,iwuoha2022protests,aidonojie2022legality,dambo2022nigeria}. The circumstances that led to the emergence of the \#EndSARS protest, and the impact of the mass action on policing and law enforcement in Nigeria have been examined in \cite{ojedokun2021mass}. While the focus in \cite{ochi2021effect} is on how the protest affects the Nigerian economy,  the work of \cite{aidonojie2022legality} is geared towards ascertaining the reasons behind the \#EndSARS protest in the hope of proffering some useful solutions. From the legal point of view, the work of \cite{iwuoha2022protests} examines how the police action (SARS unit) implicates Nigeria and the impact on democracy concerning civil and human rights. The study points to the lack of mutual trust between the government and the citizens. For online activism, Nigerians in the diaspora have played a crucial role in influencing social reforms \cite{olabode2016veterans}. This is in line with recent findings \cite{dambo2021office,abimbade2022millennial}. The online activities and characterisation of the protest have been examined to understand how the movement is promoted. The strategies used by the protesters include calling-out celebrities, unfollowing political leaders, sharing stories, fundraising and hacktivism \cite{abimbade2022millennial}. Also, the role of influencers within the movement has been analysed in a related study. It was found that users relied more on foreign media than local media for coverage while Nigerians in the diaspora promote information propagation about the protest \cite{dambo2021office}. The top 10 influencers are scored based on the connection of each user to disconnected users within the network structure \cite{dambo2021office}. This construct results in Nigeria’s president and Lagos state governor (and other media houses) being ranked high on the list. We focus on a different approach to identifying top influencers based on their online activity vis-a-vis the protest.

\section{Methodology}
\label{sec:methodology}
To answer the questions raised in Section~\ref{sec:introduction}, our approach is centred around the following (1) collecting \#EndSARS-related data from Twitter (2) performing a preliminary analysis of the data (3) identifying relevant stakeholders, performing trend, discourse and engagement analyses. 
We describe these processes in this section.

\subsection{Data Collection} 
This study utilises data from Twitter, which offers an essential mixture of diverse user groups that are suitable for research purposes. Through the platform's application programming interface (API), we developed a Python application to retrieve relevant tweets based on the terms in Table~\ref{tab:data-collection} for analysis. A \textit{tweet object} is a complex data object consisting of many descriptive fields that enables the extraction of various features related to a post on Twitter. With the aid of the API, numerous fields can be pulled out from the tweet object. Accordingly, we leverage the API to retrieve both historical and live tweets related to the \#EndSARS protest from 2017 to 2021. Table~\ref{tab:data-collection} shows a summary of the terms used for collection. In line with Twitter's regulation\footnote{\url{https://developer.twitter.com/en/developer-terms/display-requirements}}, we ensured that best practices have been followed in collecting and using the data for our research purposes. Table~\ref{tab:data-summary} shows basic statistics about the collected data. 

   \begin{table}[t]
        \small
        \caption{Hashtags used in collecting the research data from Twitter} 
        \label{tab:data-collection}
        \begin{tabular}{p{2.15cm} p{11cm}}
        \hline
        \textbf{Tweets}   & \textit{based on keywords and hashtags related to the protest} \\
        \textbf{Description}    & collection of \#EndSARS-related tweets via Twitter's API \\
        \textbf{Hashtags}   & 
                \#EndSARS (using a variation of the hashtag), police brutality, stop the killing, Nigeria Police, SARS Nigeria,
                Lekki bridge,\#endpolicebrutality \\ \hline
        \end{tabular}
    \end{table} 

\subsection{Preliminary Analysis} 
Because the data collection was conducted over time, we performed some longitudinal and exploratory analyses to explore the data. 

     \begin{table}[t]
              \footnotesize
              \caption{A summary of basic statistics in the collected datasets.}
              \label{tab:data-summary}
              \begin{tabular}{l>{\raggedright}p{3.0cm}cclp{3cm}} \hline
                & Data Kind &  Size  & Mean Values& Description  \\ \hline 
                \multirow{4}{*}{\rotatebox{90}{\textbf{Dataset}}} 
                & Raw Tweets & 3550 & 23  & average number of tokens in the tweet \\
                & Retweeted Tweets & 3422 & 463 & average retweet count \\
                & Replied Tweets & 1317 & 0.3 & average replies in the collection \\
                & Quoted Tweets & 1107 & 0.14 & average number of quoted tweets \\ \hline
              \hline
            \end{tabular}
            \normalsize
        \end{table}

\subsubsection{Activity Trend}
\begin{figure}[!ht]
    \centering
        \begin{subfigure}[t]{.4\textwidth}
            \subcaption{2012 - 2019 Period}
            \centering
           \includegraphics[width=\textwidth]{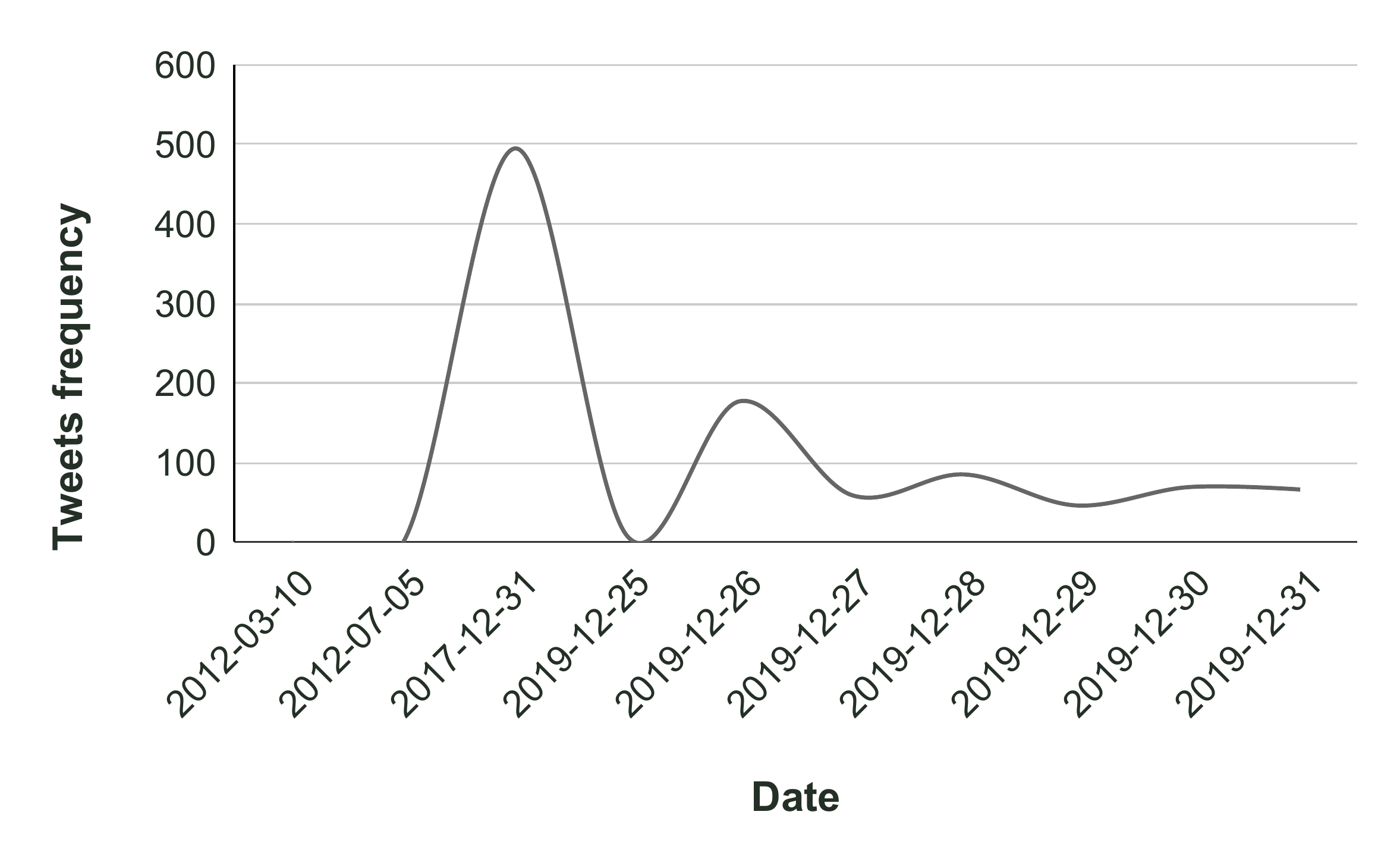}
            \label{fig:2012-2019-TRates}
        \end{subfigure}
        \hfill
        \begin{subfigure}[t]{.4\textwidth}
            \centering
            \subcaption{January to June 2020}
           \includegraphics[width=\textwidth]{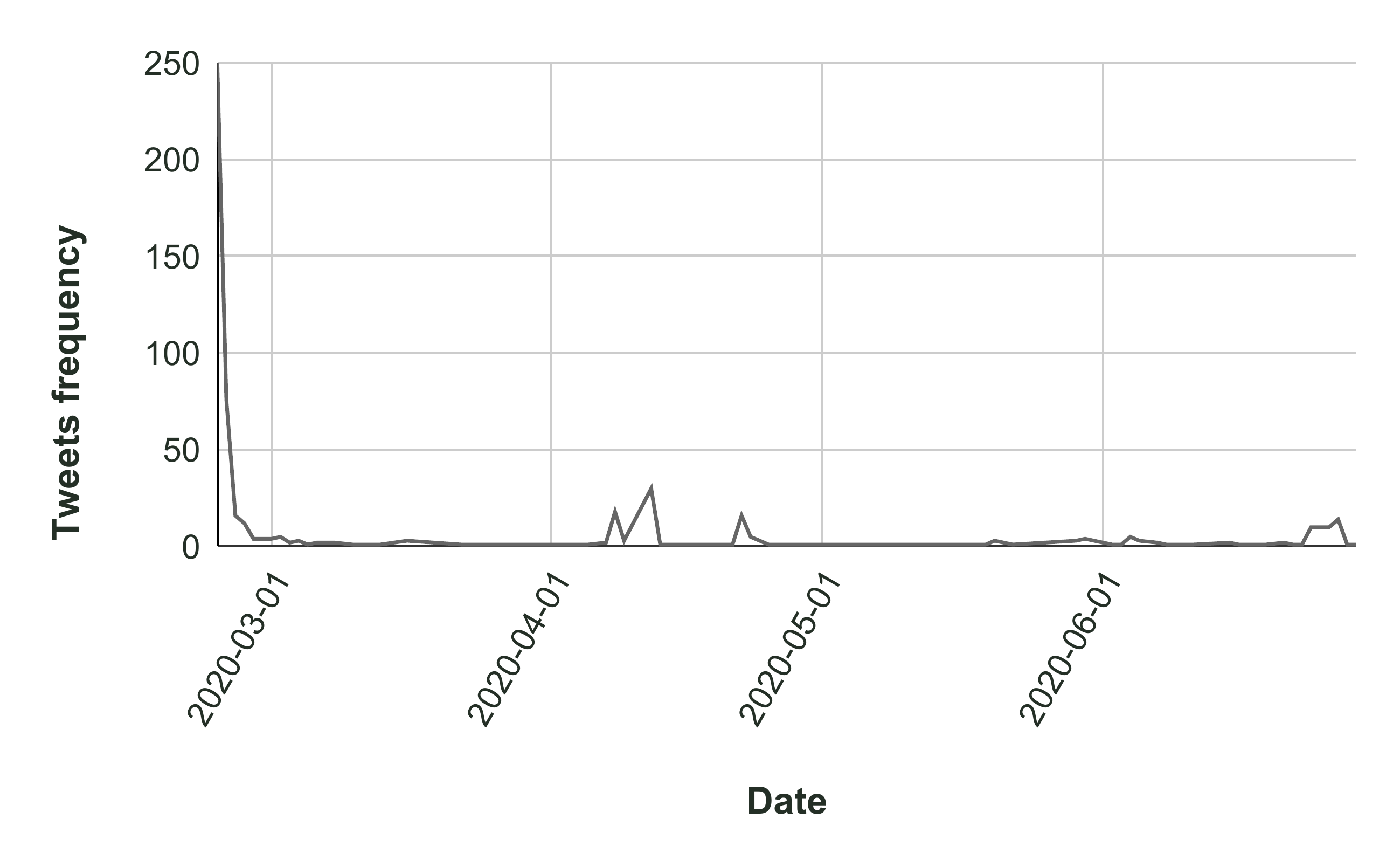}
            \label{fig:2020A-TRates}
        \end{subfigure}  
        \hfill
        \begin{subfigure}[t]{.4\textwidth}
            \centering
            \subcaption{July to December 2020}
            \includegraphics[width=\textwidth]{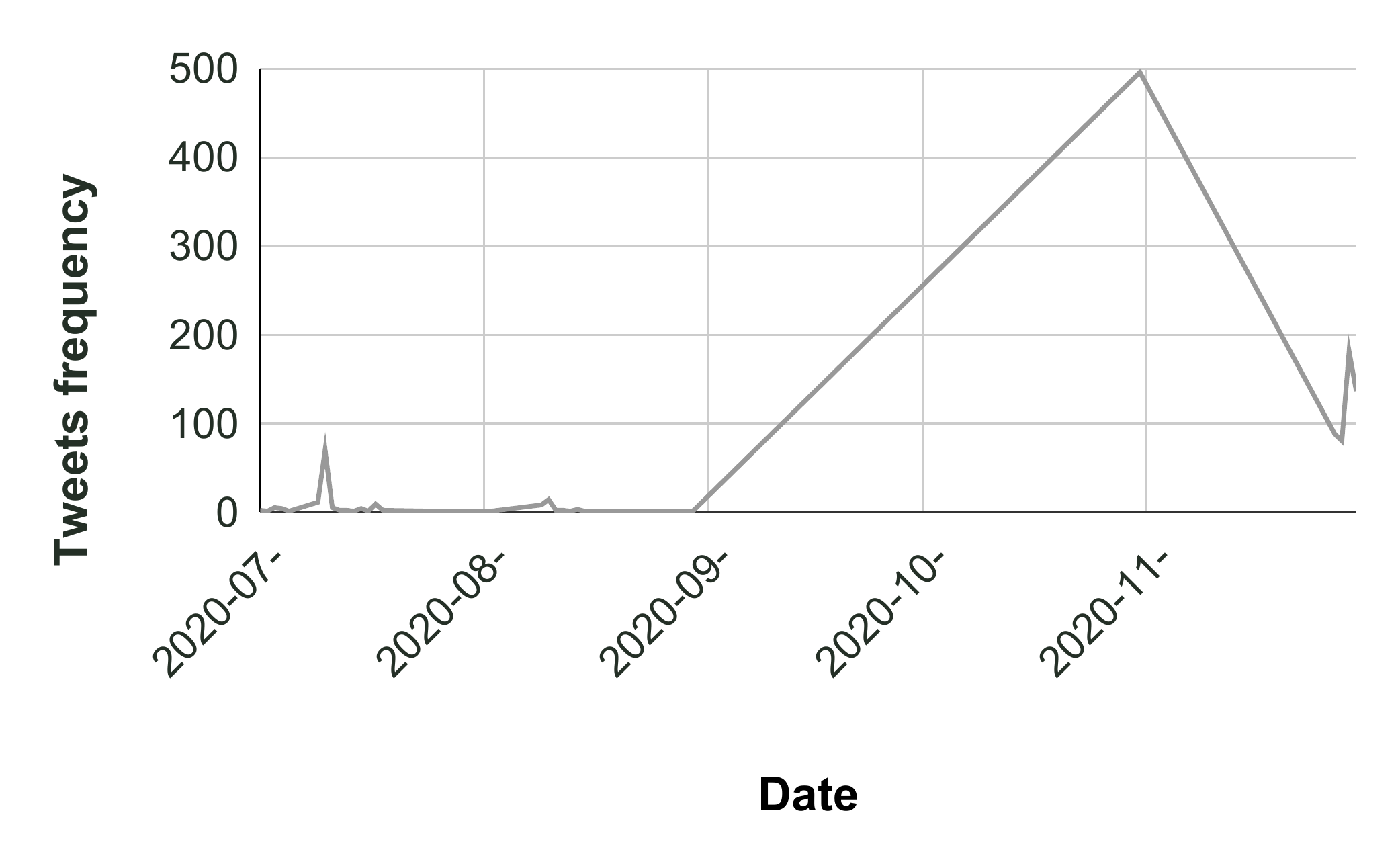}
            \label{fig:2020B-TRates}
        \end{subfigure} 
            \hfill
        \begin{subfigure}[t]{.5\textwidth}
            \centering
                \subcaption{Year 2021}            \includegraphics[width=\textwidth]{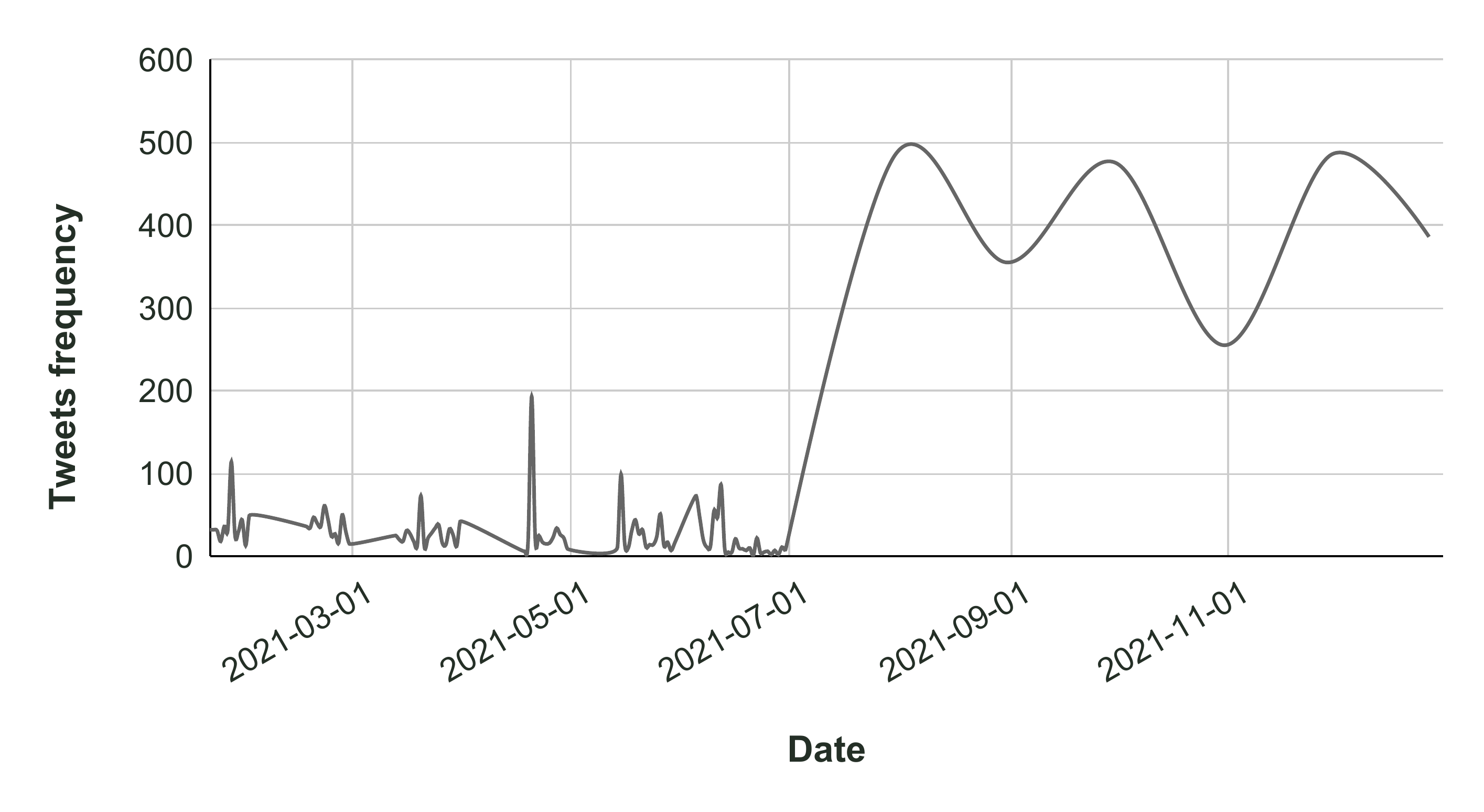}
            \label{fig:2021-TRates}
        \end{subfigure}
        \caption{Frequency of tweets related to the protest from 2017-2022 showing (a) the beginning of the protest discussion on Twitter covering 2012-2019; the spike in activity can be observed towards the end of 2017 (b) tweets about the \#EndSARS protest that were generated in 2020; and (c) tweets about the \#EndSARS protest covering 2021 to early 2022. The sub-figure (d) shows the devices mostly used by the users discussing the protest themes on Twitter.}
        \label{fig:tweetovertime}
    \end{figure}

The collection period covers 2017-2021, which corresponds to the period when discussion about the protest theme started. In Figure~\ref{fig:relevant-tweets-about-sars}, sub-figure~\ref{fig:FirstENDSARSTweet} appeared to be one of the earliest tweets to bring the SARS issue for open discussion on Twitter. During the early days of the protest, there were a handful of users discussing the issue. However, as the issue is debated over time, the engagement flared up, especially during the height of the protest (2020 - 2021), which coincides with a period when a video about a police shooting went viral on social media platforms. 
Figure~\ref{fig:tweetovertime} depicts how the online discourse on \#EndSARS varies over time. Sub-figures \ref{fig:2012-2019-TRates} to \ref{fig:2021-TRates} depict the rate of tweet generation over time. The spike in sub-figure \ref{fig:2012-2019-TRates} coincides with the shooting incident alluded to earlier (see Figure~\ref{fig:police-shooting}). Sometime in 2020, news about the alleged killing by SARS officers was shared on social media platforms, and the incidence is attributed to the observed peak in Figure~\ref{fig:2020A-TRates}. However, the veracity of the information was denied by the Government and the individual responsible was apprehended. 
Moreover, the spikes in activity in Figure~\ref{fig:2012-2019-TRates} can be attributed to self-reported encounters with the police or related traumatic experiences with the SARS officers. As time passed by, the protest recorded low online participation between January to August 2021. As shown in Figure~\ref{fig:2021-TRates}, towards the end of the same year numerous protesters increased tweeting in memory of the October 2020 incident and protest-related themes. 


\paragraph{Offline Protest}
The \#EndSARS protest happened concurrently both in offline and online modes. From Figure~\ref{fig:tweetovertime} it can be observed that since the first mention of the protest theme in 2017, the activity trend has been decreasing until late 2020 when the tragic shooting incident involving SARS officers that led to the loss of life \footnote{\url{https://www.aljazeera.com/news/2020/10/22/timeline-on-nigeria-unrest}} happened. This led to a wider protest across the country. Figure~\ref{fig:map} depicts the proportion of the \#EndSARS protest according to the 36 states in Nigeria, including the Federal Capital Territory (FCT) Abuja. It can be observed from the Figure that the most active promoters of the online protest are confined to a few locations, notably Lagos and Abuja. As the protest keeps raging, it draws the attention of various national and international stakeholders. Thousands of Nigerians across the nation, especially in Lagos and Abuja, poured onto the streets protesting over the activities of the SARS unit. This led to another encounter with the police leading to some organisations expressing concerns over the handling of the incident. For instance, Amnesty International (AI-Nigeria) cautioned the authorities over the handling of the incidence (see sub-figure \ref{fig:amnesty}). Moreover, both the African Union Commission chairman and the UN High Commissioner for Human Rights have called for normalcy to be restored. 

\section{Result and Discussion}
\label{sec:result-and-discussion}
In this section, we present relevant results and discussions about the main findings of the research. 

\subsection{Stakeholders in the online activism} 
\begin{figure}[!ht]
        \centering{
            \includegraphics[width=1.0\linewidth,height=10cm]{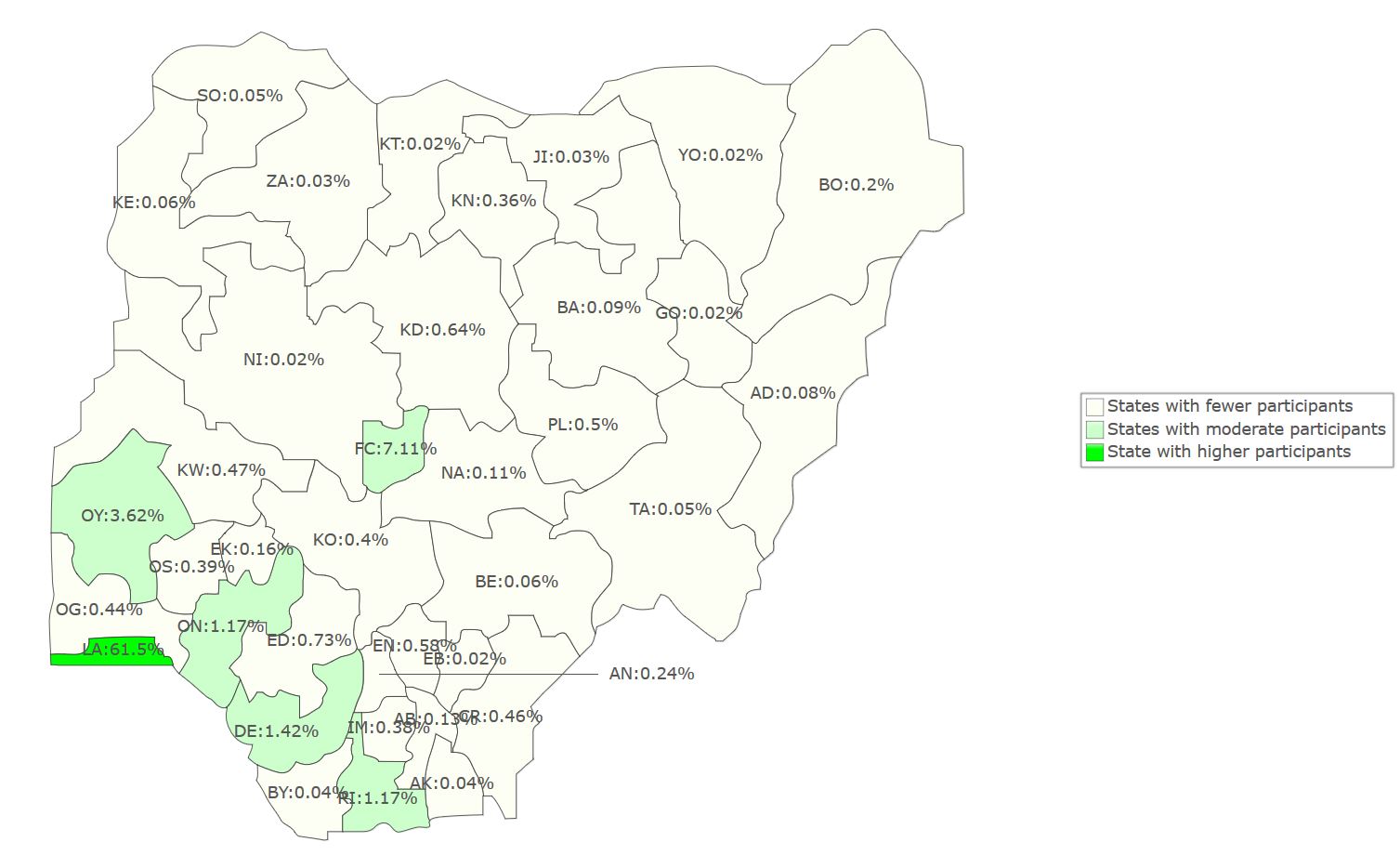}
            \caption{Proportion of the \#EndSARS online protest across the 36 states in Nigeria, including the FCT, Abuja.}
            \label{fig:map}
              }
    \end{figure}

\begin{figure}[!ht]
        \centering{
        \includegraphics[width=.7\linewidth,height=8cm]{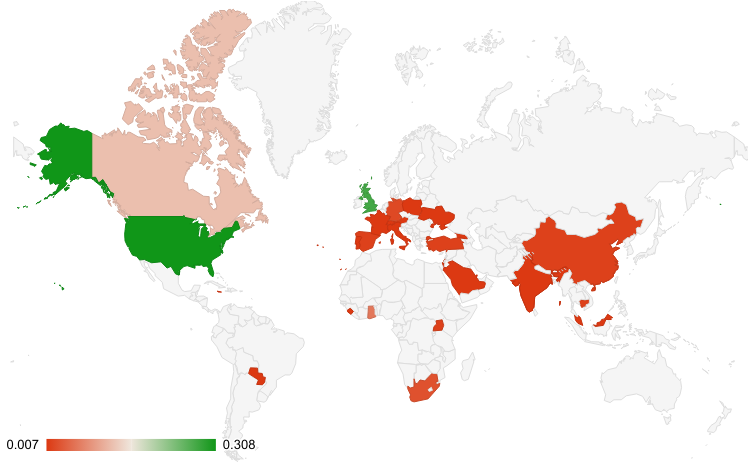}
            \caption{Proportion of the \#EndSARS online protest across the globe. There is a high proportion of engagement with the protest themes in the United States of America, United Kingdom and Canada. We can see some moderate engagements from South Africa, Ghana, Uganda, Liberia, Paraguay, Saudi Arabia, India, China, Indonesia, Malaysia, Cambodia, Turkey, and most of Western Europe Countries.}
            \label{fig:world-map}
              }
    \end{figure}

Twitter offers unique features to understand when and where an event (tweet) happened. 
The dataset we examined in this study contains information about the user's location, including the sources or devices used by the online protesters. 
Although we can not ascertain the reliability of the user's reported data due to inaccuracies or false information provided by users, the objective is to analyse how the location and device information can be leveraged to understand the stakeholders involved in the online protest. Of interest is to understand the role of national and international (the diaspora community) in the protest. 
\paragraph{Location} information about the users can inform whether the influential stakeholders in the protest are residents or overseas. 
We acknowledge that the location feature on Twitter may not be an accurate indicator of the location, but will be useful in shading some insights and informing the analysis. 
As mentioned earlier, Figure~\ref{fig:map} illustrates the level of participation in the protest across the 36 states, including the Federal Capital Territory (FCT Abuja), of the federation. The protest is more prominent in Lagos 61.5\%, FCT Abuja 7.11\%, Oyo 3.6\%, Delta 1.42\%, Ondo 1.17\%, and Rivers 1.17\% states. 
Thus, the high percentage of participation from Lagos state is in line with media report of Lagos being the hub of the \#EndSARS protest\footnote{\url{https://www.washingtonpost.com/outlook/2020/10/25/roots-endsars-protests-nigeria/},}\footnote{\url{https://www.aljazeera.com/news/2020/10/22/timeline-on-nigeria-unrest}}. As seen in Figure~\ref{fig:world-map}, the diaspora community was quite active during the online protest. 

\paragraph{Device}  
The type of device used by a user could reveal how mobile the user is. We could infer how individuals roam from one point to another to join in online protests. Suppose the device type is \textit{desktop} computer, then the user will be considered not mobile, and could point to an organisation with dedicated personnel to keep the online protest alive. Because there are numerous devices used by users in the data collection, we place the devices into 3 broad clusters consisting of third-party sources 2\%, mobile devices 88\%, and Twitter for web apps 10\%, see Figure~\ref{fig:devicesUsed}. The high proportion of mobile devices is suggestive of high mobility among the protesters and enables them to simultaneously participate in both offline and online protests. 

\subsection{Protest Themes and Public Perception} 
\begin{figure}[tbh]
        \centering{
        \includegraphics[width=\linewidth,height=10cm]{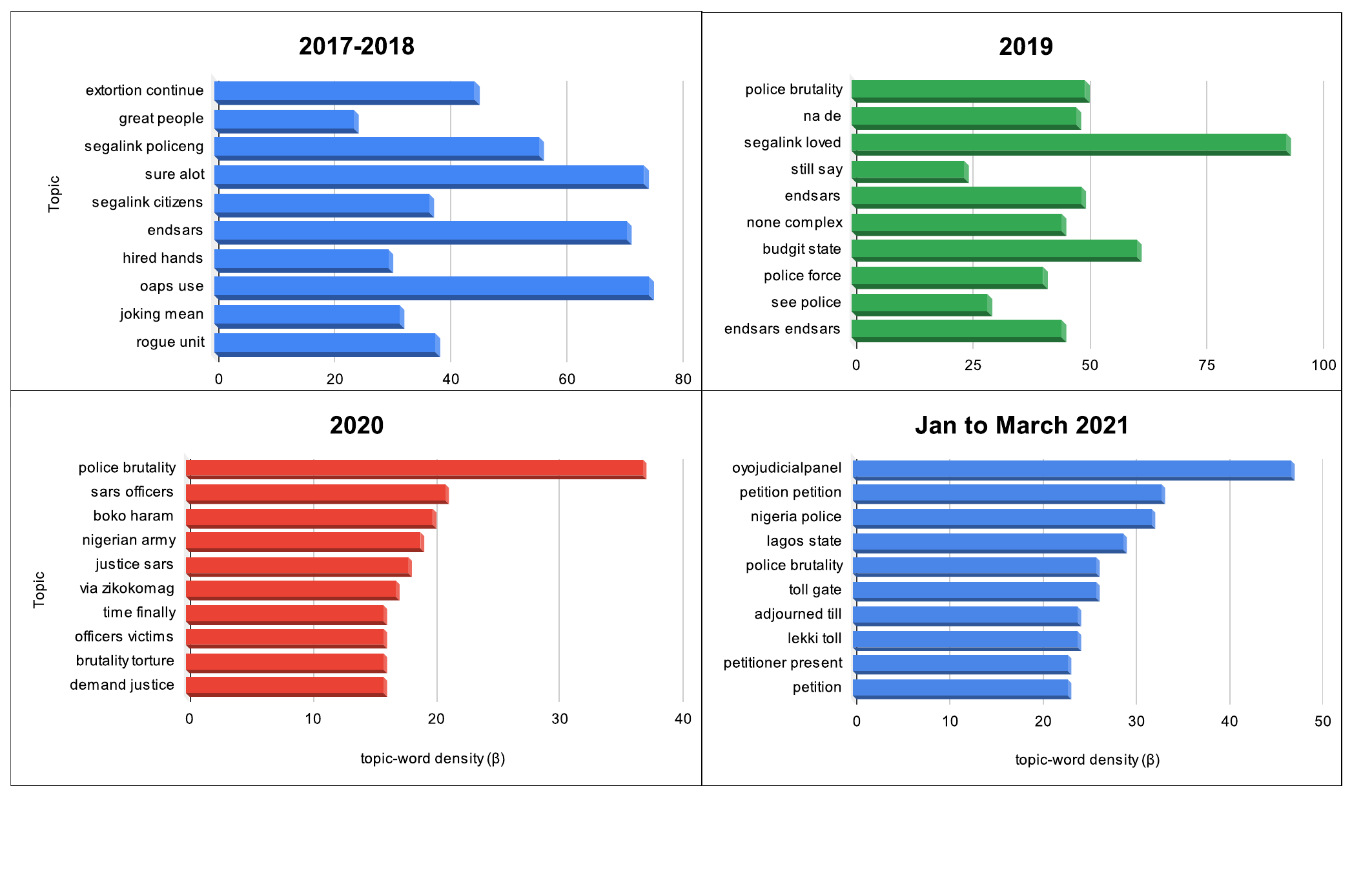}
        \caption{An overview of the main topical issues related to the protest from 2017 to March 2021. From 2017 to early 2020 the discussion topics are centred around the brutality of the SARS and advocating for support to the \#EndSARS campaign.}
        \label{fig:Topics1}
          }
    \end{figure}

\begin{figure}[tbh]
        \centering{
        \includegraphics[width=\linewidth,height=10cm]{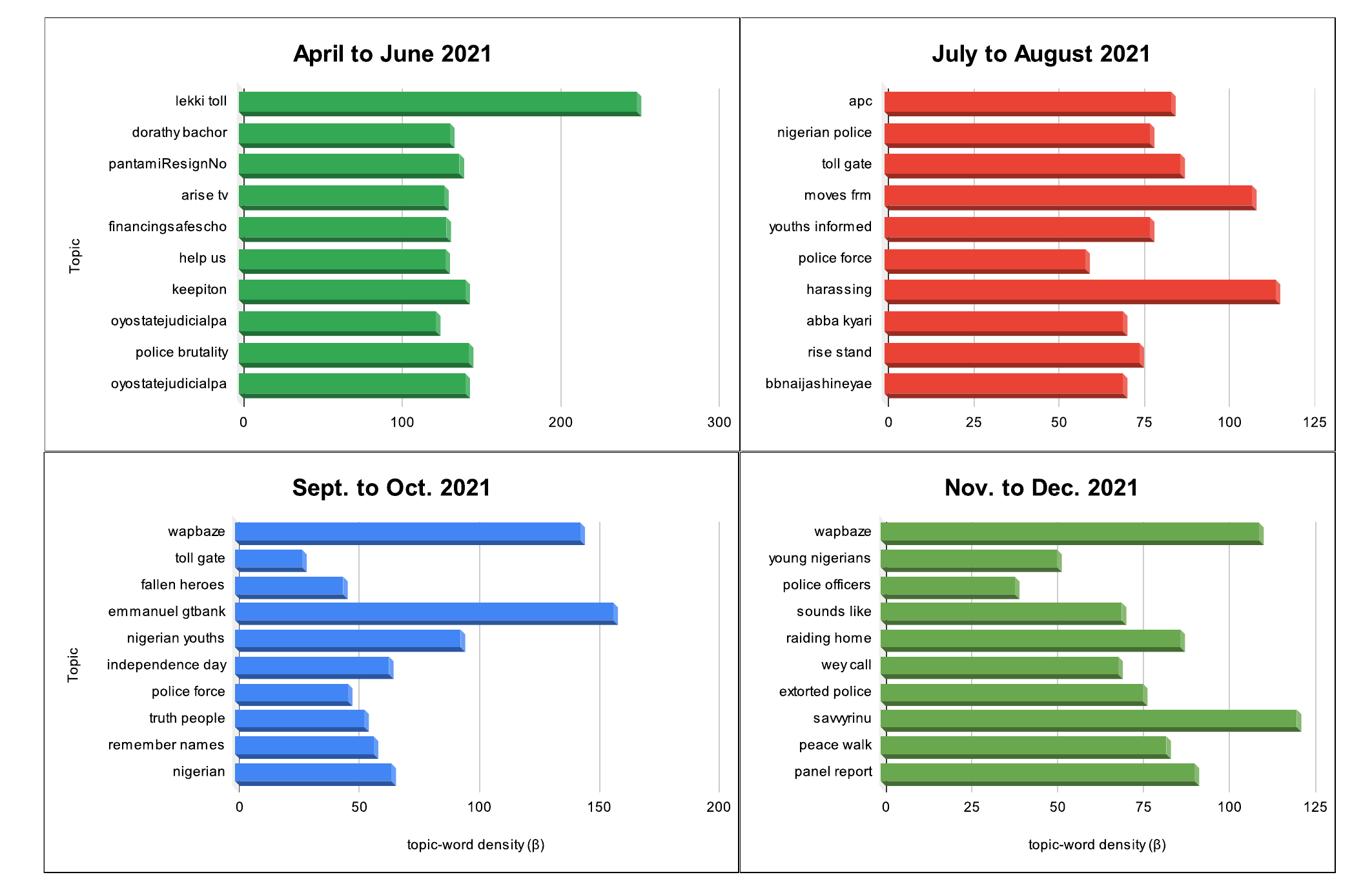}
        \caption{An overview of the main topical issues related to protest from April 2021 to December 2021. This period witnessed a high volume of activity and discussion about the protest. During this period, there is an inclusion of political movements intertwined with the protest theme.}
        \label{fig:Topics2}
          }
    \end{figure}

\paragraph{Protest Themes}
It is vital to understand the kind of topics associated with the \#EndSARS protest on Twitter. To achieve that, we perform topic modelling on the content of the raw tweets (see Table~\ref{tab:data-summary}) using the Latent Dirichlet Allocation (LDA) algorithm~\cite{blei2003latent} and the Continuous Dirichlet Allocation (CLDA) algorithm ~\cite{wang2012continuous} for a better understanding of the topics in the data. Before proceeding with the topic analysis, the tweets collection is subjected to a cleaning process to get rid of unwanted content often associated with data from Twitter \cite{inuwa2018detection}. The preprocessing step involves tokenisation, stopwords removal and text formatting. For normalisation, we remove \textit{urls, user mentions, emojis} from the collection. Figure~\ref{fig:Topics1} shows some of the relevant topics obtained from applying the LDA algorithm. For readability and ease of interpretation, we decompose the data into eight (8) different segments. For each segment, we limit our analysis to the ten (10) top topics to understand the meaning and context of the discussion topics. Figures~\ref{fig:Topics1} and ~\ref{fig:Topics2} show the respective topics under each segment. 
        
\begin{figure}[tbh]
            \centering{
            \includegraphics[width=0.6\linewidth,height=7cm]{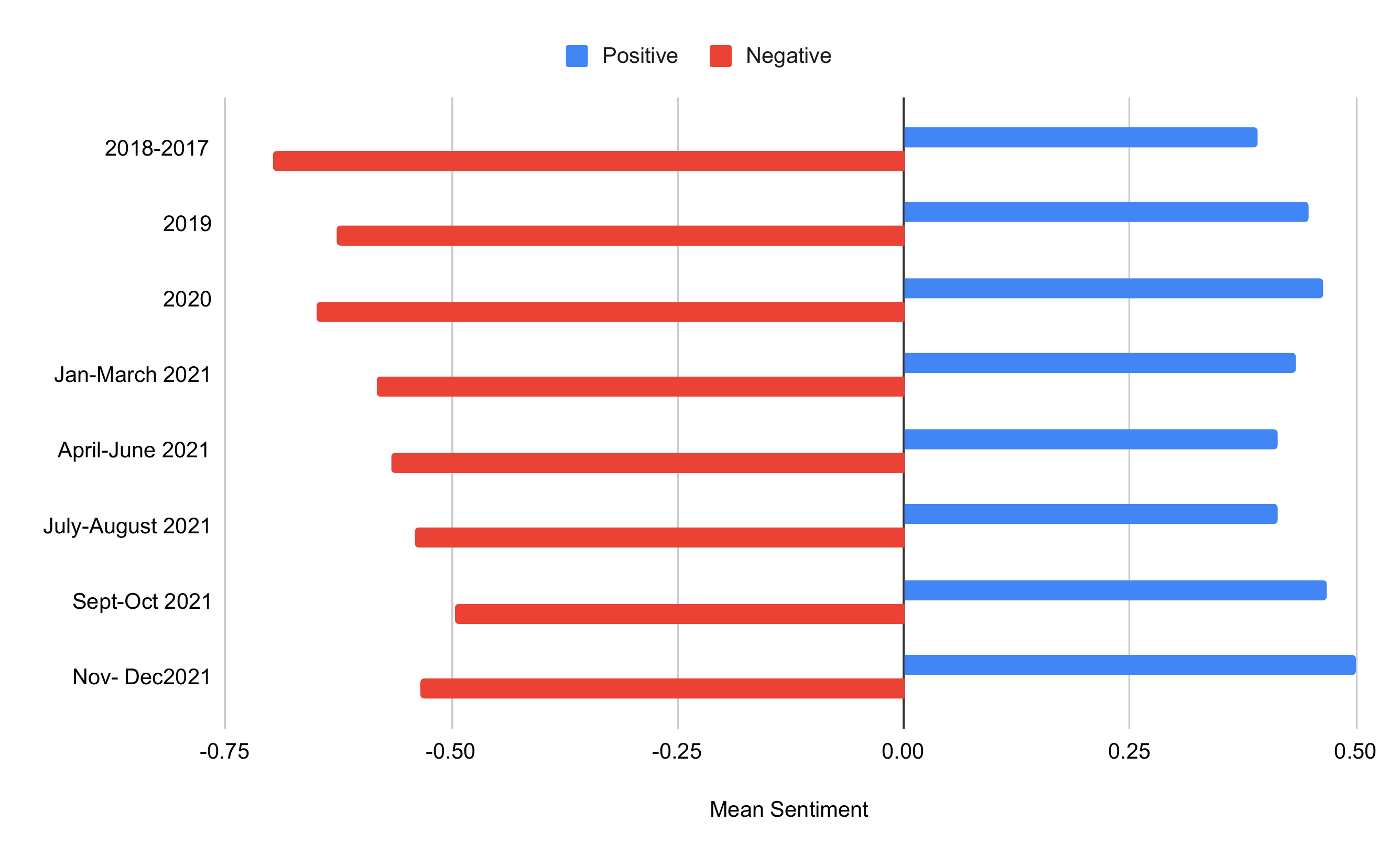}
            \caption{The polarity of sentiment associated with the \#EndSARS topics from 2017 to 2021. While the positive sentiment is mostly associated with tweets garnering public support for the movement, the negative sentiment is mainly about dissatisfaction with the activities of the SARS unit.}
            \label{fig:TopicsSentiment}
              }
    \end{figure}

\paragraph{Public Perception} 
Alongside the topical analysis, we are interested in knowing the nature or sentiment expressed in the online discourse on \#EndSARS. To accomplish that, we identify the most prominent topics during the period and pull out the associated tweets from our collection to analyse the sentiments or public perception of the protest's themes. To understand the intensity or polarity of the public opinions about the protest, we leverage the off-the-shelf VARDER~\cite{hutto2014vader} tool\footnote{a sentiment analysis tool for determining the sentiment associated with social media texts} to obtain the sentiment (as positive, neutral or negative) of each tweet. Figure~\ref{fig:TopicsSentiment} shows the result of the sentiment analysis consisting of positive and negative tweets about the movement. The negative tweets are mainly expressing the brutal actions of the SARS police while the positive tweets are mostly associated with rallying support for the \#EndSARS campaign. According to Figure~\ref{fig:TopicsSentiment}, the intensity of the negative sentiment in 2017-2018 is the highest. Noting how people tend to respond quicker to negative emotion~\cite{landmann2020being,bessi2016social}, this could be suggestive of using some emotional narratives or negative sentiment to generate some momentum and attract public attention towards the campaign. 
    
\subsection{Engagement Analysis} 
\begin{figure}[tbh]
        \centering{
            \includegraphics[width=1.0\linewidth,height=7cm]{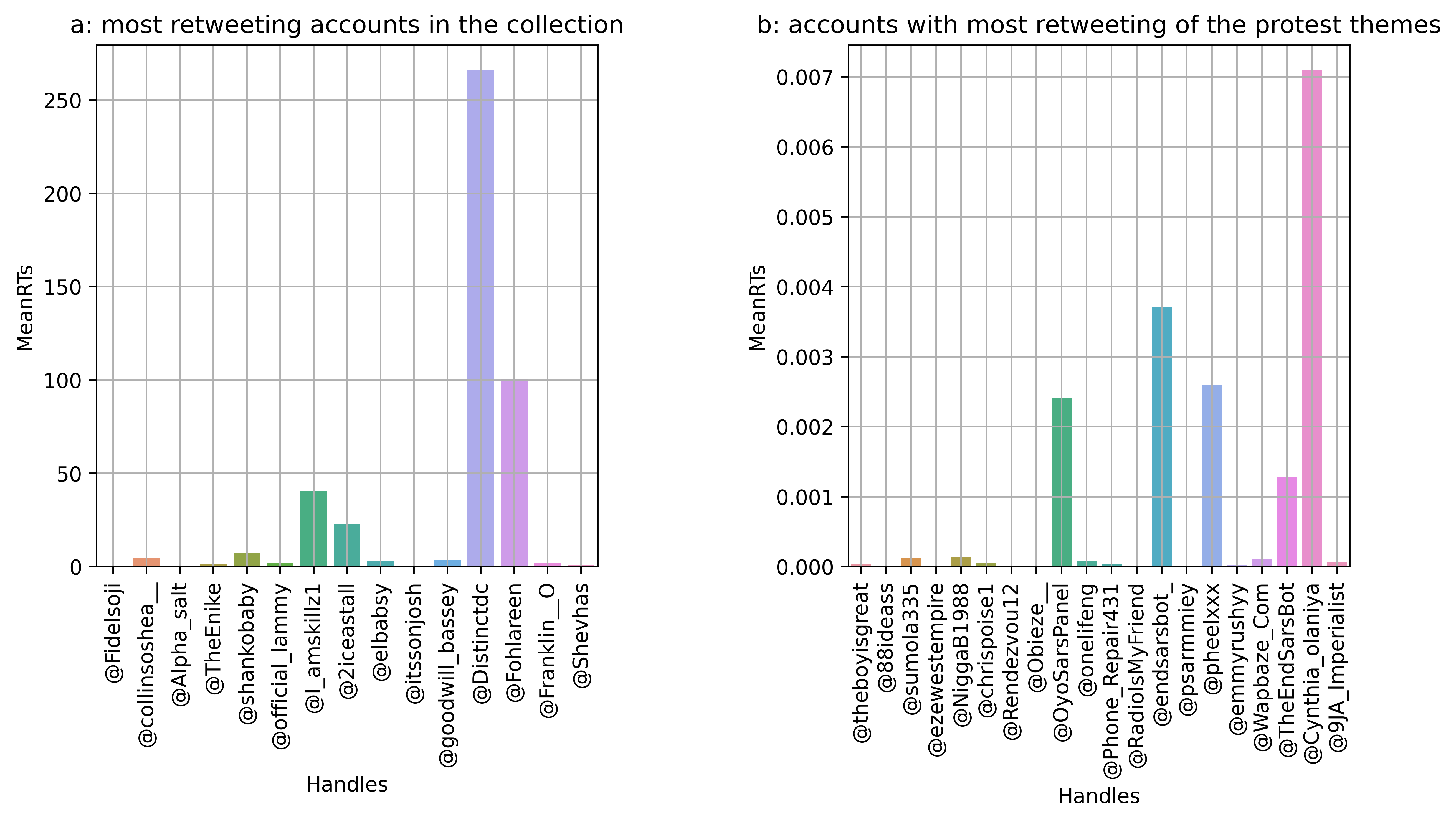}
            \caption{The level of engagement according to (a) accounts with the highest relative retweeting action in the overall collection (b) relative retweets in accounts with the highest engagement with the protest themes.}
            \label{fig:engagement-retweets}
              }
    \end{figure}

\begin{figure}[tbh]
        \centering{
            \includegraphics[width=1.0\linewidth,height=7cm]{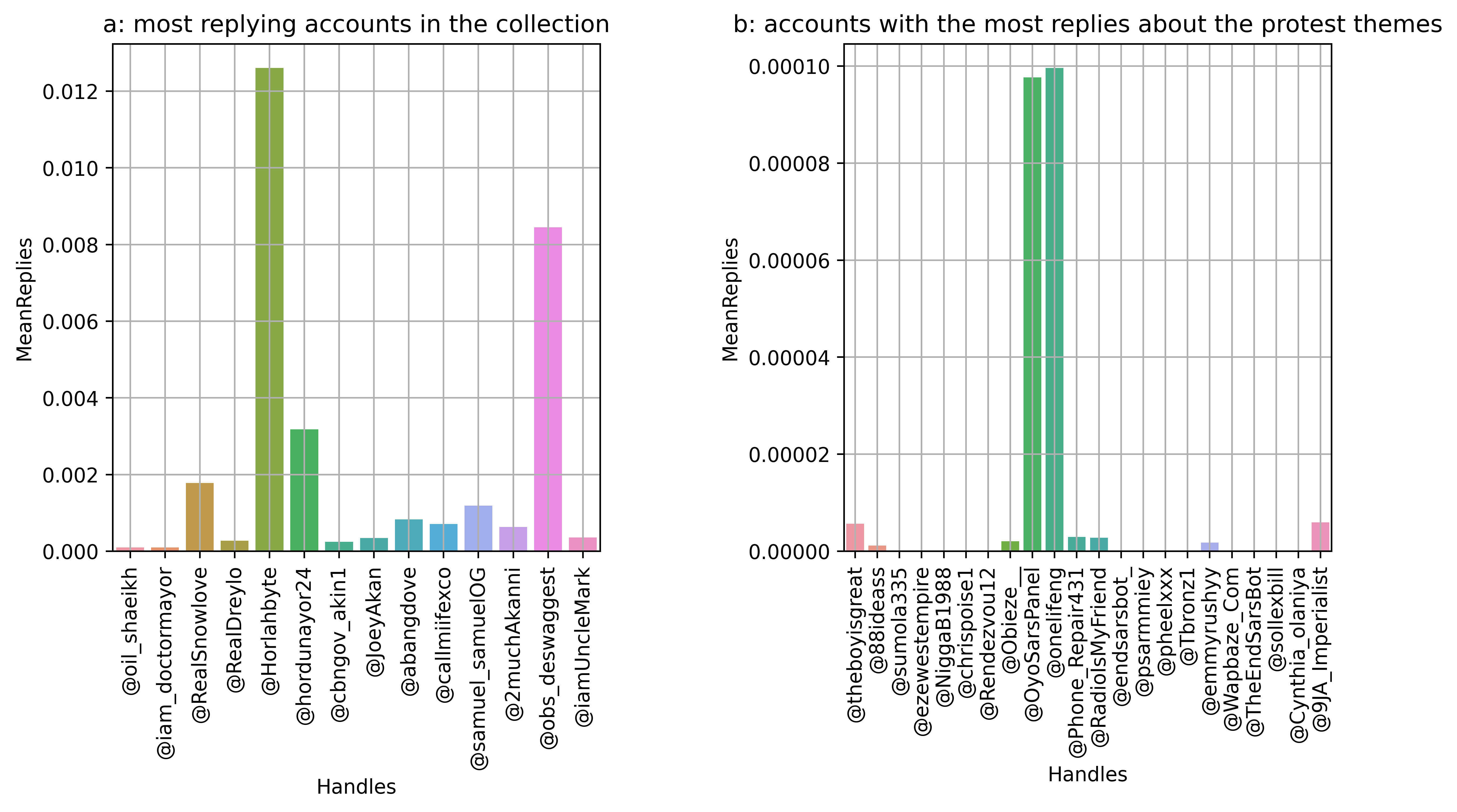}
            \caption{The level of engagement according to (a) accounts with the highest relative replying action in the overall collection (b) relative replies in accounts with the highest engagement with the protest themes.} 
            \label{fig:engagement-replies}
              }
    \end{figure}

There are many ways to examine how users engage with each other on Twitter. What constitutes engagement and how to measure it on Twitter has been analysed in previous studies \cite{inuwa2018effect,inuwa2018lexical}. 
We can capture engagement levels depending on how often a given account performs the actions of retweeting (RT), tweet liking (favourite count), replying and user mention. In this study, we focus on measuring engagement intensity concerning online activities such as the frequency of tweet generation, retweets, and replies. For the engagement analysis, we categorise an account holder as influential if the user has tweeted or posted at least 10 tweets about the \#EndSARS protest. This is to limit the number of irrelevant or insufficient tweets. Figures~\ref{fig:engagement-retweets} and ~\ref{fig:engagement-replies} show the highest engaging users based on relative retweet and reply actions. Figures~\ref{fig:engagement-tweets} show the tweet count of the most active users in the overall data collection and the collection with the highest engagement with the protest's themes. The figures also point to the possibility of using social bots to promote the online movement noting the explicit 'bot' usage in the account name. The accounts with the highest engagement with the protest themes are not influential in the literal sense of being a celebrity or known influential user. For instance, one of the most active accounts in terms of retweeting action is the \textit{@endsarsbot\_} that retweeted 429 tweets from 2021-07-31 to 2021-12-27. The main task of the account is limited to retweeting and calling out to some Twitter users as a way of attracting attention and promoting the protest. Another active account is the \textit{@psarmmiey} with 105 retweets from 2021-07-31 to 2021-12-27. Interestingly, the activity period of both \textit{@endsarsbot\_} and \textit{@psarmiey} are the same (from 2021-07-31 to 2021-12-27) and their action is limited to retweeting as a strategy to mobilise for the protest. Similarly, the \textit{@TheEndSarsBot} retweeted 38 tweets from 2021-11-30 to 2021-12-27. 

\subsection{From \#EndSARS to Demand for Change} 
In addition to the themes and public perception of the protest, it is crucial to determine whether the protest is purely driven by the need to end the SARS unit or not. Through the topic modelling, we were able to identify relevant tweets from the collected data based on the detected themes, see Figure~\ref{fig:Topics2}. This is required in order to make sense of the topics in detail.
According to Figures~\ref{fig:Topics1} and ~\ref{fig:Topics2}, it appears that from 2017-2018, the major topics were geared towards generating some momentum for the \#EndSARS movement. Some examples during this period include: 
        \begin{itemize}
            \item[-] \textbf{'extortion continue'}
            \begin{itemize}
                \item \textit{"This kind of extortion has to end, we will continue to demand \#ENDSARS. Kindly retweet so people would be aware of the kind of atrocity committed by @PoliceNG"}
            \end{itemize}
            \item[-] \textbf{'sure alot'}
            \begin{itemize}
                \item \textit{" ....kindly RT this to advice/warn our guys to be careful tonight as I am sure alot of SARs will be on the road tonight."}
            \end{itemize}
            \item[-] \textbf{'OAPs use'}
            \begin{itemize}
                \item \textit{"I will implore our OAPs to use their position to enlighten the public and not deceive them on the SARS matter. There is nothing to be politically correct about in a life and death matter. It is the same common man’s kids that is endangered. \#EndSARS.}
            \end{itemize} 
        \end{itemize} 
The last topic (\textit{'OPAs use'}), relates to the tweets calling for 'On-air Personalities (OAPs)' to support and enlighten the public about the activities of the SARS unit. Other online topical issues during the protest period include \textit{police force} and \textit{police brutality} calling for the reform of the Nigeria Police Force, which is attributed to the \textit{ Lekki toll gate} in Lagos incidence that resulted in the loss of lives and casualties in an attempt by the police to disperse protesters\footnote{https://www.bbc.co.uk/news/world-africa-59300011}. 
Although the majority of the online discourse on the protest is on ending the SARS unit, we observe the emergence of different themes concerning the hashtags used and the associated content. These hashtags are about social reforms and questioning the style of governance. An example of such hashtags with a political undertone is the \textit{\#endbadgovernanceinnigeria} alluding to regime change (see Figure~\ref{fig:hashtags-calling-for-change}). This and similar hashtags have surfaced alongside the protest theme on Twitter. Also, there exists hashtags and related tweets accusing the 
government in power of supporting \textit{boko haram}, and demanding for the president to resign. Overall, the \#EndSARS discussions in 2021 seem to revolve around the demand for justice for the protesters, ending bad governance and change of government. 

\section{Conclusion}
\label{sec:conclusion}
The Special Anti-Robbery Squad (SARS) is a special police unit under the Nigeria Police Force tasked with the responsibility of fighting violent crimes. Over the years, the unit has been accused of various human rights abuse across the nation leading to one of the major protests in the country. Due to the demographics of the protesters, the protest was quite popular on various online social networks. Using the popular \textit{\#EndSARS} hashtag on Twitter to draw attention to human rights violations labelled against the SARS unit, the protest received huge attention within and outside the country. In this study, we critically examined the evolution of the protest on Twitter, the nature of engagement with the protest themes, the factors influencing the protest and public perceptions about the online movement.  
We have examined how social media users interact with the \#EndSARS movement and how online communities respond to the protest. Based on examining public perception, we found that the \#EndSARS-related hashtags have been used to report ill-treatment by the SARS unit and to rally support for the protest. Needless to say, the themes in the online discourse are heavily on ending SARS and vicarious experience, however, there are topics around reforming the police and demand for regime change. Injecting a political undertone could lead to setbacks in social activism as it will likely deviate from the main objective of the movement or provoke counter protest. 
In line with previous findings \cite{abimbade2022millennial}, the following mobilisation strategies of calling-out celebrities, and sharing direct stories and vicarious experiences have been used to promote the protest. Based on the explicit 'bot' suffix, there is some evidence that suggests the use of automated accounts to promote the course of the protest. 
Through location and device analysis, we studied the degree of participation in the protest and how the protesters engage with the \#EndSARS themes. Over 70\% of the protest is confined within a few states in Nigeria. The diaspora communities also lent their voices to the issue. The most active users are not those with high followership. The majority of the protesters utilised mobile devices, accounting for 88\% to mobilise and report on the protest. Our future study will explore the possibility of mapping out how the online protest translated into physical violence or confrontation. Also, we will examine how information, including misinformation, propagates within the network of the protesters. 


\newpage

\appendix
\section*{Appendix}
    \begin{figure}[tbh]
        \centering
            \begin{subfigure}{.4\textwidth}
                \centering
               \includegraphics[width=.5\textwidth]{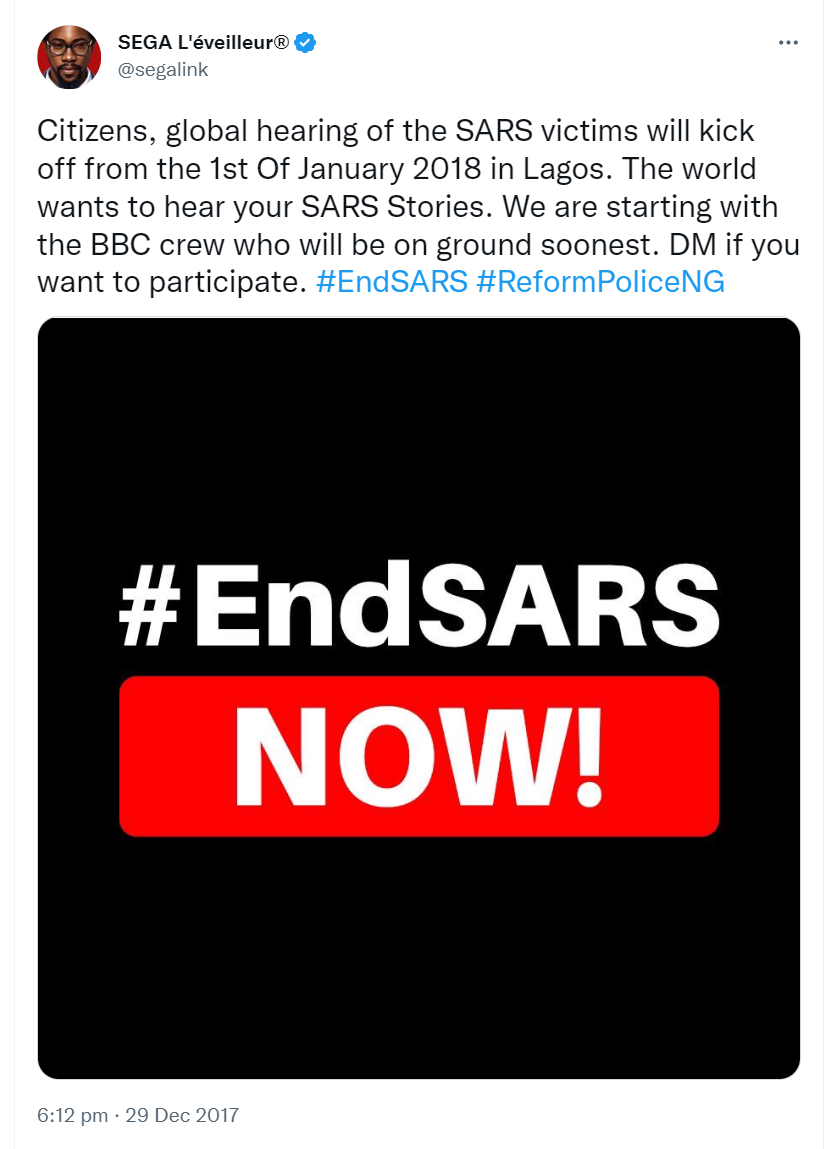}
               \subcaption{One of the earliest tweets about the \#ENDSARS movement.}
                \label{fig:FirstENDSARSTweet}
            \end{subfigure}
            \begin{subfigure}{.4\textwidth}
               \centering
               \includegraphics[width=.5\textwidth]{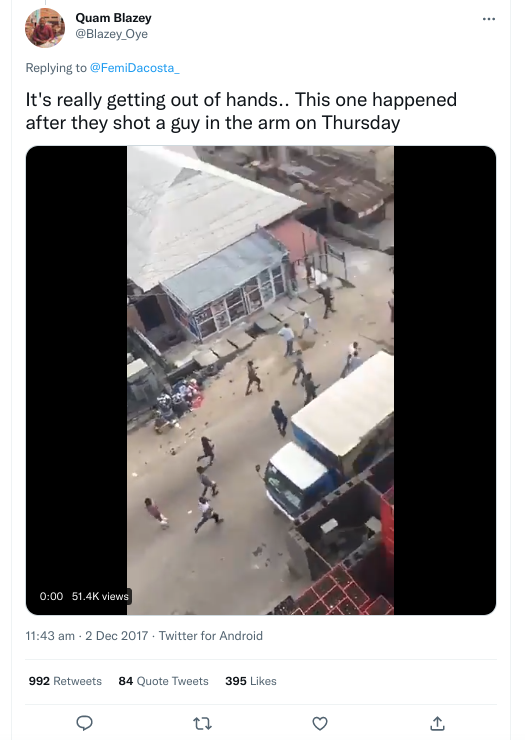}
               \subcaption{A tweet with video of police shooting in 2017. This has provoked wider protest\footnote{\url{https://twitter.com/Blazey_Oye/status/936923719825367040}}}
                \label{fig:police-shooting}
            \end{subfigure}
            \begin{subfigure}{.4\textwidth}
               \centering
               \includegraphics[width=.5\textwidth]{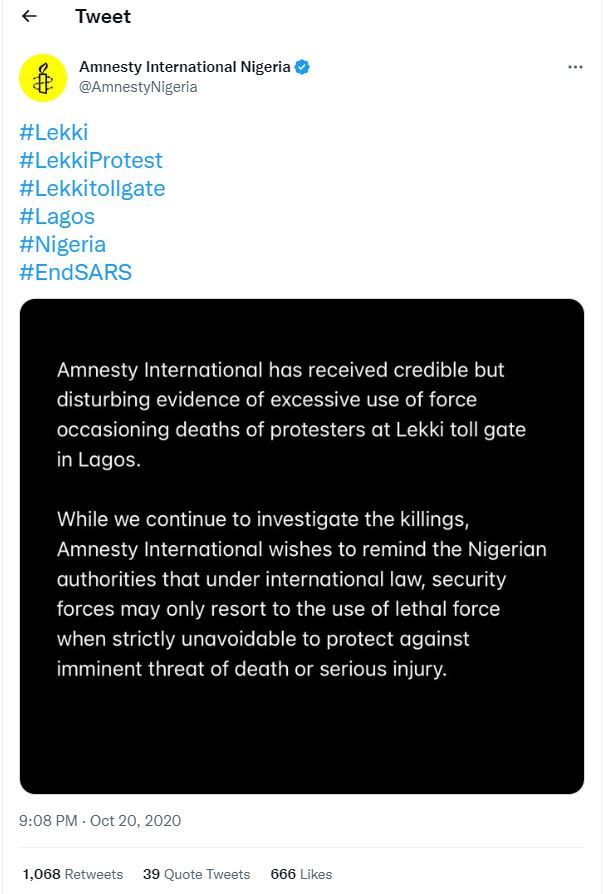}
               \subcaption{A sample tweet from Amnesty International, Nigeria about handling of the protest by the authority.}
                \label{fig:amnesty}
            \end{subfigure}
            \begin{subfigure}{.4\textwidth}
               \centering
               \includegraphics[width=.6\textwidth]{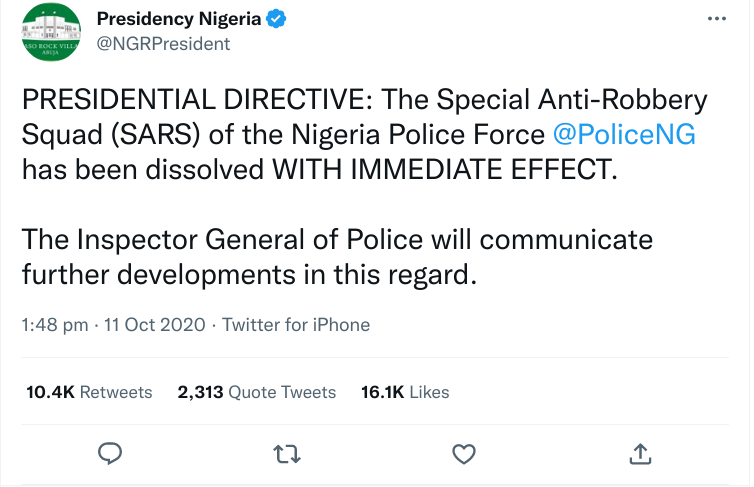}
                \subcaption{A directive from the Government of Nigeria banning SARS.}
                \label{fig:SARS-ban}
            \end{subfigure}
             \caption{Relevant tweets about the \#EndSARS movement}
            \label{fig:relevant-tweets-about-sars}
            \hfill
        \end{figure}


    \begin{figure}[tb!]
        \centering
            \begin{subfigure}{.4\textwidth}
                \centering
               \includegraphics[width=\textwidth]{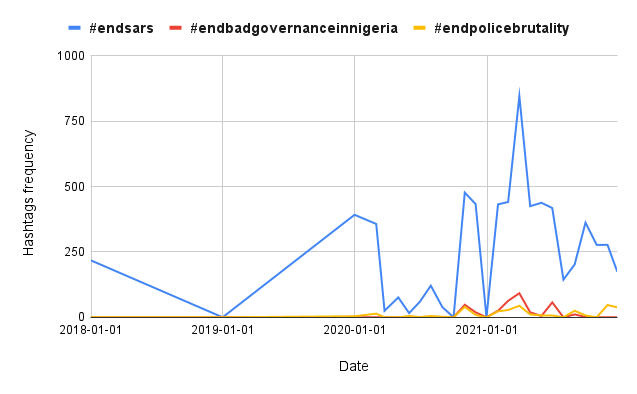}
               \subcaption{Some relevant hashtags}
                \label{fig:hashtags-calling-for-change}
            \end{subfigure}
            \begin{subfigure}{.4\textwidth}
               \centering
               \includegraphics[width=\textwidth]{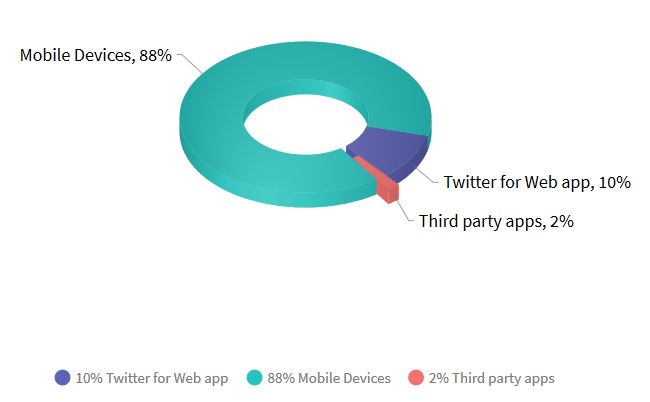}
               
               \subcaption{Devices used by the online protesters.}
                \label{fig:devicesUsed}
            \end{subfigure}
             \caption{The (a) main \#EndSARS hashtag alongside hashtags calling for change in the style of governance by the protesters (b) main devices used by the protesters.}
            \label{fig:hashtags-calling-for-change-and-devices}
            \hfill
        \end{figure}

    \begin{figure}[ht!]
        \centering{
            \includegraphics[width=0.9\linewidth,height=6cm]{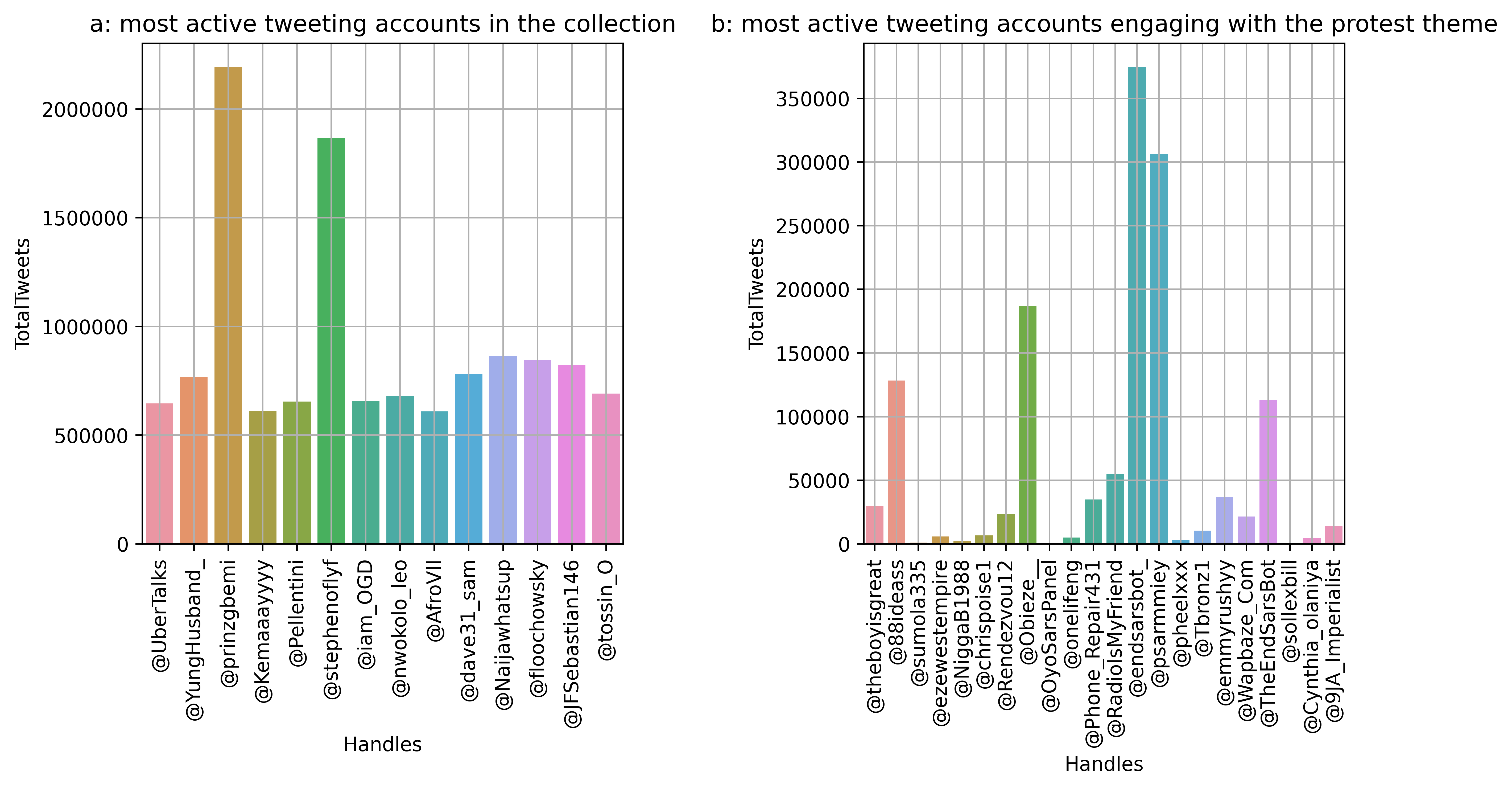}
            \caption{The level of engagement according to (a) accounts with the highest number of tweets in the overall collection (b) accounts with the highest engagement with the protest themes.}
            \label{fig:engagement-tweets}
              }
    \end{figure}

  \newpage

\bibliographystyle{unsrt}
\bibliography{end-sars}

\begin{thebibliography}{10}

\bibitem{amnesty-sars2020}
Nigeria Amnesty~International.
\newblock Nigeria: Time to end impunity - torture and other violations by
  special anti-robbery squad (sars).
\newblock \url{https://www.amnesty.org/en/documents/afr44/9505/2020/en/}, 2020.

\bibitem{brandl1994global}
Steven~G Brandl, James Frank, Robert~E Worden, and Timothy~S Bynum.
\newblock Global and specific attitudes toward the police: Disentangling the
  relationship.
\newblock {\em Justice quarterly}, 11(1):119--134, 1994.

\bibitem{weitzer2004race}
Ronald Weitzer and Steven~A Tuch.
\newblock Race and perceptions of police misconduct.
\newblock {\em Social problems}, 51(3):305--325, 2004.

\bibitem{warren2011perceptions}
Patricia~Y Warren.
\newblock Perceptions of police disrespect during vehicle stops: A race-based
  analysis.
\newblock {\em Crime \& Delinquency}, 57(3):356--376, 2011.

\bibitem{kane2002social}
Robert~J Kane.
\newblock The social ecology of police misconduct.
\newblock {\em Criminology}, 40(4):867--896, 2002.

\bibitem{mastrofski2002police}
Stephen~D Mastrofski, Michael~D Reisig, and John~D McCluskey.
\newblock Police disrespect toward the public: An encounter-based analysis.
\newblock {\em Criminology}, 40(3):519--552, 2002.

\bibitem{jones2013march}
William~P Jones.
\newblock {\em The March on Washington: Jobs, freedom, and the forgotten
  history of civil rights}.
\newblock WW Norton \& Company, 2013.

\bibitem{ince2017social}
Jelani Ince, Fabio Rojas, and Clayton~A Davis.
\newblock The social media response to black lives matter: How twitter users
  interact with black lives matter through hashtag use.
\newblock {\em Ethnic and racial studies}, 40(11):1814--1830, 2017.

\bibitem{buggs2017dating}
Shantel~Gabrieal Buggs.
\newblock Dating in the time of\# blacklivesmatter: Exploring mixed-race
  women’s discourses of race and racism.
\newblock {\em Sociology of Race and Ethnicity}, 3(4):538--551, 2017.

\bibitem{byrd2017vitality}
W~Carson Byrd, Keon~L Gilbert, and Joseph~B Richardson~Jr.
\newblock The vitality of social media for establishing a research agenda on
  black lives and the movement.
\newblock {\em Ethnic and Racial Studies}, 40(11):1872--1881, 2017.

\bibitem{haffner2019place}
Matthew Haffner.
\newblock A place-based analysis of\# blacklivesmatter and counter-protest
  content on twitter.
\newblock {\em GeoJournal}, 84(5):1257--1280, 2019.

\bibitem{ohia2020covid}
Chinenyenwa Ohia and Mobolaji~Modinat Salawu.
\newblock Covid-19 pandemic and civil unrests in africa: implication of
  recent\# endsars protests for increased community transmission in nigeria.
\newblock {\em The Pan African Medical Journal}, 37(Suppl 1), 2020.

\bibitem{uwazuruike2020endsars}
Allwell~Raphael Uwazuruike.
\newblock \# endsars: the movement against police brutality in nigeria.
\newblock {\em Harvard Human Rights Journal}, 2020.

\bibitem{ajisafe2021impacts}
Dickson Ajisafe, Tinuade~Adekunbi Ojo, and Margaret Monyani.
\newblock The impacts of social media on the\# endsars\# youth protests in
  nigeria.
\newblock In {\em Proceedings of the ICTeSSH 2021 Conference. Proceedings of
  the ICTeSSH 2021 conference, Virtual conference. https://doi.
  org/10.21428/7a45813f. 638ef816}, 2021.

\bibitem{dambo2021office}
Tamar~Haruna Dambo, Metin Ersoy, Ahmad~Muhammad Auwal, Victor~Oluwafemi
  Olorunsola, and Mehmet~Bahri Saydam.
\newblock Office of the citizen: a qualitative analysis of twitter activity
  during the lekki shooting in nigeria’s\# endsars protests.
\newblock {\em Information, Communication \& Society}, pages 1--18, 2021.

\bibitem{ekoh2021role}
Prince~Chiagozie Ekoh and Elizabeth~Onyedikachi George.
\newblock The role of digital technology in the endsars protest in nigeria
  during covid-19 pandemic.
\newblock {\em Journal of human rights and social work}, 6(2):161--162, 2021.

\bibitem{iwuoha2022protests}
Victor~Chidubem Iwuoha and Ernest~Toochi Aniche.
\newblock Protests and blood on the streets: Repressive state, police brutality
  and\# endsars protest in nigeria.
\newblock {\em Security Journal}, 35(4):1102--1124, 2022.

\bibitem{aidonojie2022legality}
Paul~Atagamen Aidonojie, Oluwaseye~Oluwayomi Ikubanni, and Alade~Adeniyi
  Oyebade.
\newblock Legality of endsars protest: A quest for democracy in nigeria.
\newblock {\em Journal of Human Rights, Culture and Legal System},
  2(3):209--224, 2022.

\bibitem{dambo2022nigeria}
Tamar~Haruna Dambo, Metin Ersoy, Ahmad~Muhammad Auwal, Victor~Oluwafemi
  Olorunsola, Ayodeji Olonode, Abdulgaffar~Olawale Arikewuyo, and Ayodele
  Joseph.
\newblock Nigeria's\# endsars movement and its implication on online protests
  in africa's most populous country.
\newblock {\em Journal of Public Affairs}, 22(3):e2583, 2022.

\bibitem{amaza2020}
M~Amaza.
\newblock Heinrich boll stiftung, 2020.

\bibitem{olabode2016veterans}
Shola Olabode.
\newblock Veterans of diaspora activism: An overview of ict uses amongst
  nigerian migrant networks.
\newblock {\em The digital transformation of the public sphere}, pages
  129--148, 2016.

\bibitem{miller2000geography}
Byron~A Miller.
\newblock {\em Geography and social movements: comparing antinuclear activism
  in the Boston area}, volume~12.
\newblock U of Minnesota Press, 2000.

\bibitem{sewell2001space}
William Sewell et~al.
\newblock Space in contentious politics.
\newblock {\em Silence and voice in the study of contentious politics}, 78:18,
  2001.

\bibitem{mccaughey2003cyberactivism}
Martha McCaughey and Michael~D Ayers.
\newblock {\em Cyberactivism: Online activism in theory and practice}.
\newblock Psychology Press, 2003.

\bibitem{graham2016content}
Roderick Graham and ‘Shawn Smith.
\newblock The content of our\# characters: Black twitter as counterpublic.
\newblock {\em Sociology of Race and Ethnicity}, 2(4):433--449, 2016.

\bibitem{weitzer2005racially}
Ronald Weitzer and Steven~A Tuch.
\newblock Racially biased policing: Determinants of citizen perceptions.
\newblock {\em Social forces}, 83(3):1009--1030, 2005.

\bibitem{kennedy2006beyond}
Helen Kennedy.
\newblock Beyond anonymity, or future directions for internet identity
  research.
\newblock {\em New media \& society}, 8(6):859--876, 2006.

\bibitem{warf2018sage}
Barney Warf.
\newblock {\em The SAGE Encyclopedia of the Internet}.
\newblock Sage, 2018.

\bibitem{miller2010data}
Harvey~J Miller.
\newblock The data avalanche is here. shouldn’t we be digging?
\newblock {\em Journal of Regional Science}, 50(1):181--201, 2010.

\bibitem{haffner2018spatial}
Matthew Haffner.
\newblock A spatial analysis of non-english twitter activity in houston, tx.
\newblock {\em Transactions in GIS}, 22(4):913--929, 2018.

\bibitem{longley2015geotemporal}
Paul~A Longley, Muhammad Adnan, and Guy Lansley.
\newblock The geotemporal demographics of twitter usage.
\newblock {\em Environment and Planning A}, 47(2):465--484, 2015.

\bibitem{sheet2018social}
Fact Sheet.
\newblock Social media fact sheet. pew research center, 2022.

\bibitem{kumar2014twitter}
Shamanth Kumar, Fred Morstatter, and Huan Liu.
\newblock {\em Twitter data analytics}.
\newblock Springer, 2014.

\bibitem{ojedokun2021mass}
Usman~A Ojedokun, Yetunde~O Ogunleye, and Adeyinka~A Aderinto.
\newblock Mass mobilization for police accountability: The case of
  nigeria’s\# endsars protest.
\newblock {\em Policing: A Journal of Policy and Practice}, 15(3):1894--1903,
  2021.

\bibitem{ochi2021effect}
Ijeoma~Brigid Ochi and Kingsley~Chinonso Mark.
\newblock Effect of the endsars protest on the nigerian economy.
\newblock {\em Global Journal of Arts, Humanities and Social Sciences},
  9(3):1--15, 2021.

\bibitem{abimbade2022millennial}
Oluwadara Abimbade, Philip Olayoku, and Danielle Herro.
\newblock Millennial activism within nigerian twitterscape: From mobilization
  to social action of\# endsars protest.
\newblock {\em Social Sciences \& Humanities Open}, 6(1):100222, 2022.

\bibitem{blei2003latent}
David~M Blei, Andrew~Y Ng, and Michael~I Jordan.
\newblock Latent dirichlet allocation.
\newblock {\em Journal of machine Learning research}, 3(Jan):993--1022, 2003.

\bibitem{wang2012continuous}
Chong Wang, David Blei, and David Heckerman.
\newblock Continuous time dynamic topic models.
\newblock {\em arXiv preprint arXiv:1206.3298}, 2012.

\bibitem{inuwa2018detection}
Isa Inuwa-Dutse, Mark Liptrott, and Ioannis Korkontzelos.
\newblock Detection of spam-posting accounts on twitter.
\newblock {\em Neurocomputing}, 315:496--511, 2018.

\bibitem{hutto2014vader}
Clayton Hutto and Eric Gilbert.
\newblock Vader: A parsimonious rule-based model for sentiment analysis of
  social media text.
\newblock In {\em Proceedings of the international AAAI conference on web and
  social media}, volume~8, pages 216--225, 2014.

\bibitem{landmann2020being}
Helen Landmann and Anette Rohmann.
\newblock Being moved by protest: Collective efficacy beliefs and injustice
  appraisals enhance collective action intentions for forest protection via
  positive and negative emotions.
\newblock {\em Journal of Environmental Psychology}, 71:101491, 2020.

\bibitem{bessi2016social}
Alessandro Bessi and Emilio Ferrara.
\newblock Social bots distort the 2016 us presidential election online
  discussion.
\newblock {\em First monday}, 21(11-7), 2016.

\bibitem{inuwa2018effect}
Isa Inuwa-Dutse, Bello~Shehu Bello, Ioannis Korkontzelos, and Reiko Heckel.
\newblock The effect of engagement intensity and lexical richness in
  identifying bot accounts on twitter.
\newblock {\em IADIS International Journal on WWW/Internet}, 16(2), 2018.

\bibitem{inuwa2018lexical}
Isa Inuwa-Dutse, Bello~Shehu Bello, and Ioannis Korkontzelos.
\newblock Lexical analysis of automated accounts on twitter.
\newblock In {\em International Conferences on WWW/Internet, ICWI 2018 and
  Applied Computing 2018}, pages 75--82. IADIS Press, 2018.

\end{thebibliography}
\end{document}